\documentclass[noshowpacs,amsmath,twocolumn,
aps,prb]
{revtex4-1}

\usepackage{setspace}
\usepackage{amsmath}
\usepackage{graphicx}
\usepackage[nearskip,margin = 0pt]{subfig}
\usepackage{subfig}

\usepackage{verbatim}
\usepackage{amsfonts}
\usepackage{amssymb}
\usepackage{epstopdf} 
\usepackage{xcolor}
\usepackage{color}
\DeclareGraphicsExtensions{.pdf,.eps,.png,.jpg,.mps} 
\usepackage{ragged2e}
\usepackage{hyperref}
\hypersetup{
    colorlinks=true,
    linkcolor=blue,
    citecolor=blue,
    filecolor=blue, 
    urlcolor=blue,
}
\usepackage{mathrsfs}

\begin{document}

\title{Rhythmic soliton interactions for integrated dual-microcomb spectroscopy}

\author{Zihao Wang,$^{1}$ Yifei Wang,$^{1}$ Baoqi Shi,$^{2,3}$ Chen Shen,$^{2}$ Wei Sun,$^{2}$ Yulei Ding,$^{1}$ Changxi Yang,$^{1}$ Junqiu Liu,$^{2,4}$ and Chengying Bao$^{1,*}$ \\
$^1$State Key Laboratory of Precision Measurement Technology and Instruments, Department of Precision Instruments, Tsinghua University, Beijing 100084, China.\\
$^2$International Quantum Academy, Shenzhen 518048, China.\\
$^3$Department of Optics and Optical Engineering, University of Science and Technology of China, Hefei, Anhui 230026, China.\\
$^4$Hefei National Laboratory, University of Science and Technology of China, Hefei 230088, China.\\
Corresponding authors:  $^*$cbao@tsinghua.edu.cn 
}

\maketitle
\newcommand{\ts}{\textsuperscript}

\newcommand{\tsb}{\textsubscript}


{\bf Rotation symmetry of microresonators supports the generation of phase-locked counter-propagating (CP) solitons that can potentially miniaturize dual-comb systems. Realization of these dual-comb compatible solitons in photonic integrated circuits remains a challenge. Here, we synthesized such CP solitons in an integrated silicon nitride microresonator and observed forced soliton oscillation due to rhythmic, time-varying soliton interactions. The interactions result in seconds mutual-coherence passively. Temporal motion in the soliton streams is discerned by measuring a quadratic-scaling frequency noise peaks and an inverse quadratic-scaling microcomb sidebands. By generating a CP soliton trimer to have two synchronized solitons in one of the orbiting directions, we resolve the incapability of measuring two unsynchronized CP soliton dimer pulses by optical cross-correlation, and show CP solitons undergo complex motion trajectory. We further prove that precise dual-comb spectroscopy with an acquisition time as short as 0.6 $\mu$s is feasible using these solitons, although the temporal motion limits the dynamic range. Besides revealing soliton interactions with different group velocities, our work propels the realization of photonic integrated dual-comb spectrometers with high passive coherence.}

The concept of mapping optical frequencies into radiofreqeucnies (RFs) by multi-heterodyne beating two optical frequency combs \cite{Schiller_OL2002spectrometry} has unlocked the potential for spectroscopy \cite{Newbury_Optica2016,Vahala_Science2016}, ranging \cite{Newbury_NP2009rapid}, time/frequency transfer \cite{Newbury_Nature2023quantum} and holography \cite{Picque2021dual}. However, dual-comb measurements necessitate complex systems to ensure mutual coherence between the combs \cite{Newbury_Optica2016}, which has limited their broader applications. 
The emergence of counter-propagating (CP) soliton microcombs, created by counter-pumping a single microcavity, offers a promising solution to address this challenge \cite{Kippenberg_NP2014,Vahala_NP2017Counter,Gaeta_OL2018CP,Vahala_Science2018Range,Kippenberg_NP2018spatial,Vahala_Science2019vernier,Vahala_NC2021architecture}. In addition to their inherent advantage in miniaturization \cite{Kippenberg_Science2018Review,Diddams_Science2020optical}, CP soliton interactions via the Rayleigh backscattering endow high mutual coherence passively \cite{Vahala_NP2017Counter}. The interactions stabilize the repetition rate difference ($\delta f_r$) to an integer fraction of the counter-pumping frequency difference (i.e., $\delta f_r=\delta\nu_P/n$, where $\delta\nu_P$ is the pump frequency difference and $n$ is a comb line number with respect to the pump) \cite{Vahala_NP2017Counter}. Thus, CP solitons in this state form a pair of Vernier-like microcombs with a line pair overlapping at the $n$th line \cite{Vahala_Science2019vernier}. However, CP solitons with this Vernier frequency locking (VFL) have only been observed in tapered fiber coupled silica wedge microresonators \cite{Vahala_Science2019vernier,Vahala_NP2017Counter,Vahala_NC2021architecture}, to our knowledge. Although CP solitons have been reported for silicon nitride (Si$_3$N$_4$) microresonators, no VFL was observed as a frequency degenerate counter-pump was used and the comb line number is limited \cite{Gaeta_OL2018CP}. Therefore, generating CP solitons with VFL in Si$_3$N$_4$ microresonators can be a significant step towards full photonic integration of dual-comb systems. 


CP solitons also provide a new pathway for delving into soliton physics \cite{Vahala_NP2017Counter,Vahala_NP2021,Matsko_OL2017bose,Kippenberg_PRL2019Polychromatic}. The multi-line-pair actions in VFL can result in rhythmic soliton interactions \cite{Bao_PRAppl2023vernier}. Thus, time-varying soliton interactions in dissipative cavities can be experimentally explored, offering a unique perspective to the commonly studied time-constant interactions. Furthermore, compared to weakly guiding silica or MgF$_2$ microcavities \cite{Vahala_NP2017Counter,Vahala_NP2021,Kippenberg_PRL2019Polychromatic}, tightly confined Si$_3$N$_4$ photonic integrated circuits (PICs) offer much greater flexibility in tailoring microcavity properties \cite{Kippenberg_NC2021high}, which empowers us to examine soliton physics in an extended parameter space.

\begin{figure*}[th!]
\begin{centering}
\includegraphics[width=\linewidth]{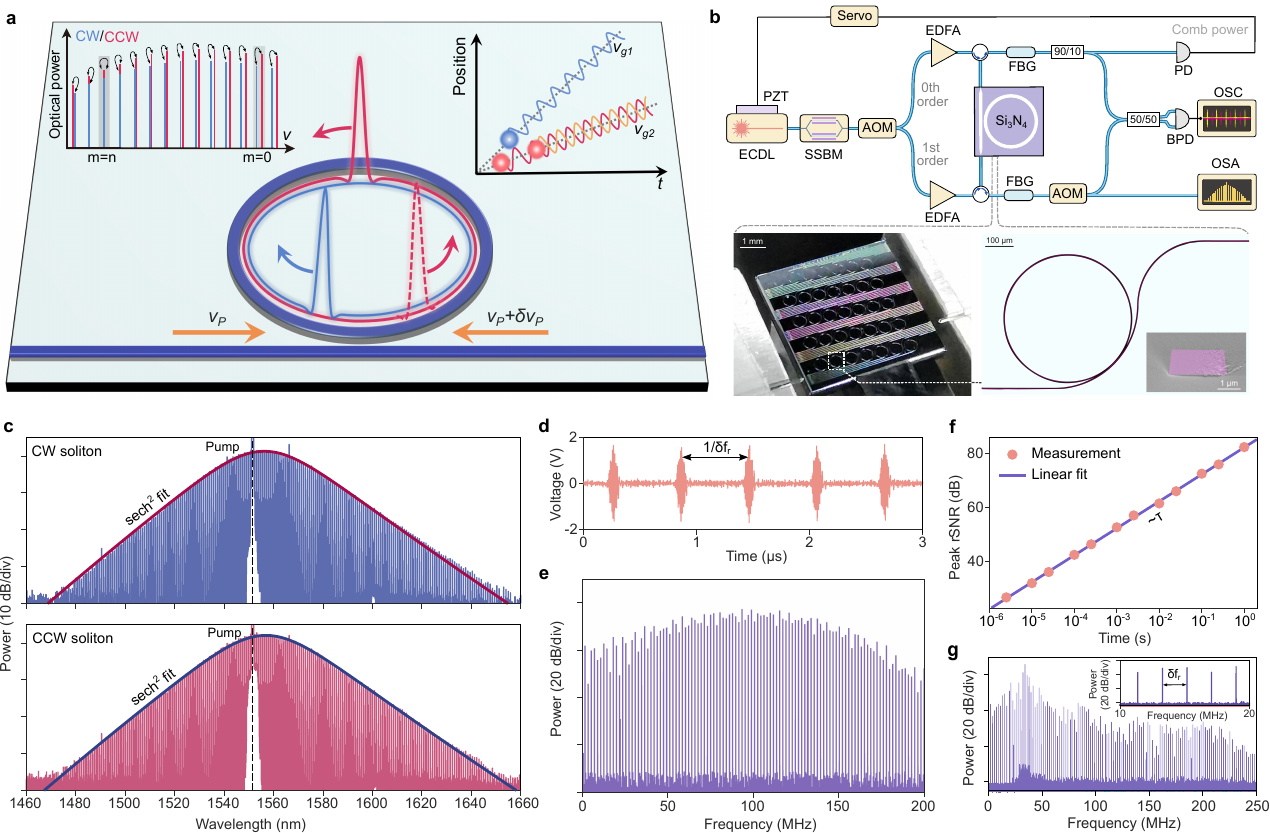}
\captionsetup{singlelinecheck=off, justification = RaggedRight}
\caption{{\bf Counter-propagating (CP) solitons in a silicon nitride microresonator.} \textbf{a,} Illustration of CP solitons in a Si$_3$N$_4$ microresonator. Due to the rhythmic soliton interactions in the Vernier frequency locking, the output CP solitons feature periodic motion in the pulse timing. \textbf{b,} Experimental setup for the generation of CP solitons. ECDL, external cavity diode laser; SSBM, single-sideband modulator; AOM, acousto-optical modulator; FBG, fiber Bragg grating; EDFA, erbium-doped fiber amplifier. The bottom panels show the picture of the Si$_3$N$_4$ photonic chip, as well as an SEM picture of the microresonator. \textbf{c,} Optical spectra of the CP soliton microcombs. \textbf{d,} Temporal interferogram of the CP solitons. \textbf{e,} Fourier transform of the interferogram of the CP soliton streams. \textbf{f,} Peak RF signal-to-noise ratio (rSNR) for the spectrum in panel \textbf{e}. The power (amplitude) rSNR scales linearly (as square-root) with the measurement time $\tau$, which suggests the mutual coherence time between the CP solitons exceeds 1 s. \textbf{g,} Photodetected electric power spectrum when selecting an individual comb line from one of the CP soliton microcombs. The spectrum comprises an array of RF lines spaced by $\delta f_r$.}
\label{Fig1}
\end{centering}
\end{figure*}

In this work, we demonstrated CP solitons with VFL in a Si$_3$N$_4$ microresonator \cite{Liu_PR2023foundry} and used them for dual-comb spectroscopy (DCS) \cite{Newbury_Optica2016}. Due to the multi-line-intractions with varying frequency spacings, the synthesized CP soliton dimers and trimers feature periodic temporal vibration (Fig. \ref{Fig1}a). Our frequency noise and balanced optical cross-correlator (BOC) based measurements \cite{Vahala_NP2021} both confirm the motion within the soliton compounds. The motion induces complex fine structures for the soliton microcombs. Despite the emergence of these VFL relevant sidebands, we show that highly coherent DCS is viable using Si$_3$N$_4$ CP solitons. However, the VFL induced soliton motion imposes limitation on the DCS measurement dynamic range. Our work adds to the fundamentals of dissipative solitons and unveils the strength and weakness of using CP solitons for PIC-based dual-comb systems.  

\begin{figure*}[th!]
\begin{centering}
\includegraphics[width=\linewidth]{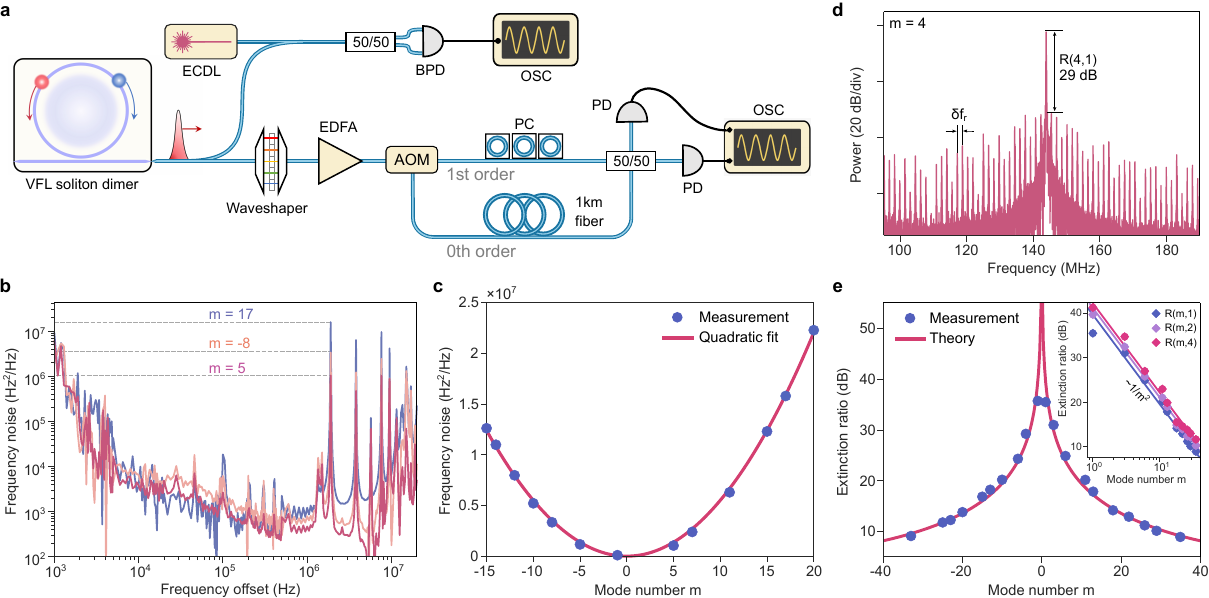}
\captionsetup{singlelinecheck=off, justification = RaggedRight}
\caption{{\bf Fine microcomb line structures due to rhythmic soliton interactions.} \textbf{a,} Experimental setup for the measurement of microcomb line frequency noise and heterodyne beat with an external cavity diode laser (ECDL) to measure the extinction ratio $R(m,N)$. \textbf{b,} Measured frequency noise spectra for three comb lines, showing strong peaks at $\delta f_r$ and its harmonics. \textbf{c,} The intensity of the first harmonic scales quadratically with the mode number $m$. \textbf{d,} Measured heterodyne beat spectrum between the $m$=4 line and the ECDL, showing multiple sidebands spaced by $\delta f_r$. \textbf{e,} Measured extinction ratio between the main line and the 1st VFL sidebands, showing an inverse quadratic trend as the theory predicts. The inset confirms this trend in a log-log plot.}
\label{Fig2}
\end{centering}
\end{figure*}

\begin{figure*}[th!]
\begin{centering}
\includegraphics[width=\linewidth]{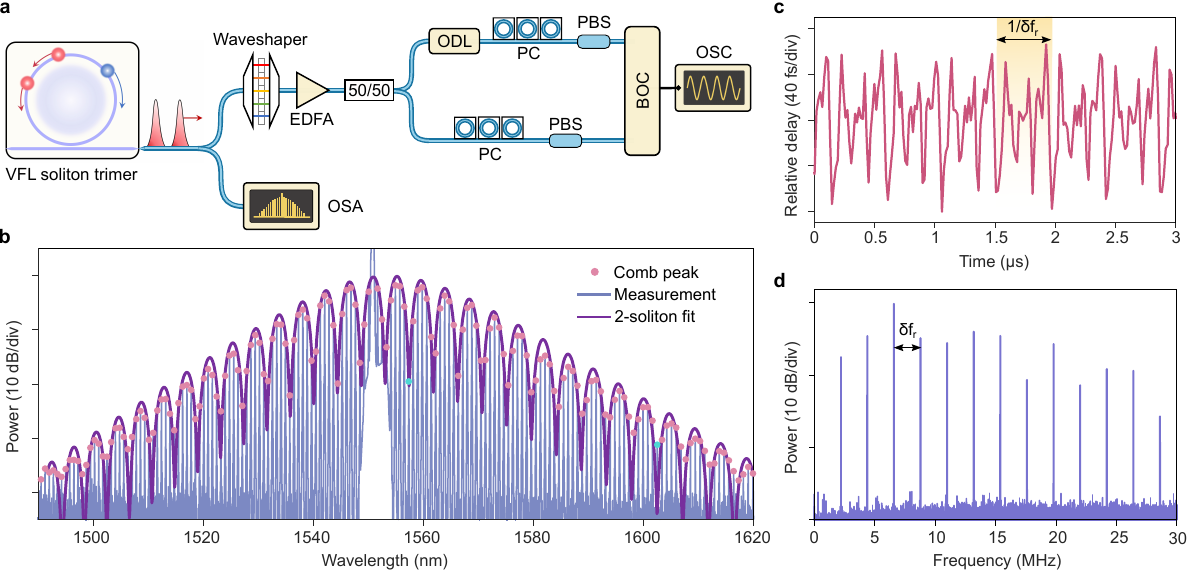}
\captionsetup{singlelinecheck=off, justification = RaggedRight}
\caption{{\bf Soliton motion within a counter-propagating (CP) soliton trimer.} \textbf{a,} Experimental setup to measure the soliton motion by a balanced optical cross-correlator (BOC). ODL, optical delay line; PBS, polarization beam splitter. \textbf{b,} Measured soliton spectrum in the direction hosting two solitons. The spectral fringes show decreasing contrast with increasing frequency spacing from the pump, see the two green dots. \textbf{c,} BOC measured soliton motion between two solitons within the soliton trimer, showing a complex motion trajectory with about 140 fs peak-to-peak amplitude. \textbf{d,} Power spectrum of the BOC measured soliton motion trajectory.}
\label{Fig3}
\end{centering}
\end{figure*}

\begin{figure*}[th!]
\begin{centering}

\includegraphics[width=\linewidth]{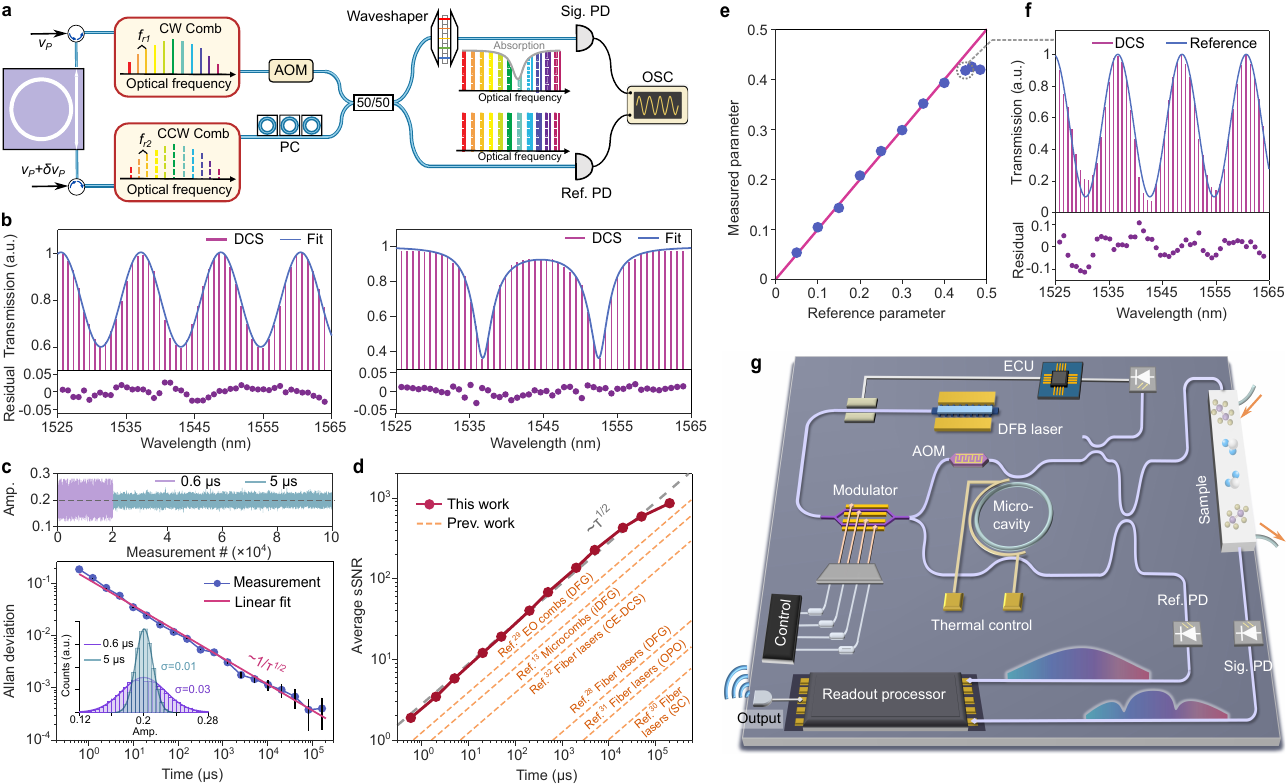}
\captionsetup{singlelinecheck=off, justification = RaggedRight}
\caption{{\bf Dual-comb spectroscopy (DCS) using counter-propagating (CP) solitons.} \textbf{a,} Illustration of the experimental setup for the CP solitons-based DCS. \textbf{b,} Measured absorption spectrum when encoding the virtual sample (a pulse shaper) with a sinusoidal or Lorentzian spectrum. \textbf{c,} Allan deviation of the sine-absorption amplitude, which scales as $1/\sqrt{\tau}$ with measurement time $\tau$. \textbf{d,} Spectroscopic signal-to-noise ratio (sSNR) of the Si$_3$N$_4$ CP soliton dual-comb spectrometer, which scales as $\sqrt{\tau}$. An average sSNR as high as 300 is reached within 10 ms, that is higher than other reports. DFG, difference frequency generation; iDFG, interleaved DFG; CE-DCS, cavity enhanced DCS; OPO, optical parametric oscillator; SC, supercontinuum. \textbf{e,} Measured sine-absorption amplitude versus applied absorption amplitude for the pulse shaper. When the absorption is high, the VFL sidebands cause the measurement to deviate from the applied absorption. \textbf{f,} Example of the distorted DCS spectrum when the applied absorption sine-amplitude is 45\% (peak-to-peak amplitude 90\%). \textbf{g,} Conceptualization of a CP soliton dual-comb spectrometer in a Si$_3$N$_4$-based photonic integrated circuit. ECU, electronic control unit.}
\label{Fig4}
\end{centering}
\end{figure*}

\noindent \textbf{Generation of Si$_3$N$_4$ CP solitons with VFL.} The experimental setup is illustrated in Fig. \ref{Fig1}b. The 100 GHz Si$_3$N$_4$ microresonator has a cross-section of 0.81$\times$2.4 $\mu$m, an intrinsic (loaded) Q-factor of 12.7 (7.8) million, and a group velocity dispersion $\beta_2$=$-$94 ps$^2$/km \cite{Liu_PR2023foundry}. A pump laser was split into two arms, with one arm frequency shifted by 55 MHz using an acousto-optical modulator (AOM). The frequency shift also contributes to reducing thermal effect, as the two pumps enter the resonance at different moments. To further mitigate thermal effects, we used the fast laser sweep technique for CP soliton generation (Fig. \ref{Fig1}b) \cite{Papp_PRL2018,liu2023mitigating}. A power lock servo was employed to ensure long term stability of the system. The generated soliton spectra are shown in Fig. \ref{Fig1}c. Since the two pumping directions have different frequency detunings from the resonance, the two solitons have a center frequency difference of $\sim$200 GHz due to the Raman response \cite{Kippenberg_PRL2016raman,Vahala_OL2016theory}. It causes the CP solitons to lose synchronization on group velocity ($v_g$) and have a nonzero $\delta f_r$. 

To test the mutual coherence of the Si$_3$N$_4$ CP solitons, we heterodyne beat them on a balanced photodetector (BPD). The interferogram shows RF pulses spaced by 1/$\delta f_r$ (Fig. \ref{Fig1}d, $\delta f_r$=1.9 MHz, VFL with $n$=29). This $\delta f_r$ is more than an order of magnitude larger than that of silica CP solitons \cite{Vahala_NC2021architecture}. Even larger $\delta f_r$ is possible by using an AOM with a higher drive frequency (i.e., a larger $\delta \nu_P$). Another AOM was inserted after the microresonator to shift one of the solitons to avoid spectral aliasing \cite{Vahala_NC2021architecture}. Fourier transform of the interferogram yields a multi-heterodyne RF comb with a high signal-to-noise ratio (rSNR, ratio between the RF lines and the averaged noise floor), see Fig. \ref{Fig1}e. The power (amplitude) rSNR scales linearly (as square root) with the averaging time $\tau$ (Fig. \ref{Fig1}f). This trend indicates the mutual coherence time between Si$_3$N$_4$ CP solitons exceeds 1 s and is limited by the measurement time only. With an averaging time of 1 s, the power (amplitude) rSNR can exceed 80 dB (40 dB). Such a high rSNR results from the high power per microcomb line and the long coherence time. The long coherence time is remarkable, as it occurs passively without any frequency locking circuits. 

As a result of multi-line-pair participation in VFL, there are optical sidebands spaced by $\delta f_r$ for each microcomb modes \cite{Bao_PRAppl2023vernier}. When filtering an individual microcomb line from one of the CP solitons and detecting it by a photodetector, multiple RF lines spaced by $\delta f_r$ is observed (Fig. \ref{Fig1}g). It unveils the role of multi-line-pair interaction that has not been experimentally investigated in all microresonator platforms yet.   

\noindent \textbf{Fine line structures in VFL soliton dimer.} Since amplitude modulation of microcomb lines also creates sidebands, Fig. \ref{Fig1}g alone cannot verify the temporal motion between output CP solitons. A microcomb can be written as $\nu_m=\nu_P+mf_r$, with $\nu_m$, $\nu_P$, $f_r$ and $m$ being the comb line frequency, pump frequency, repetition rate and line number with respect to the pump, respectively. When there is temporal motion at a rate of $\delta f_r$ for CP solitons with VFL, $f_r$ changes periodically. Consequently, the instantaneous frequency of $\nu_m$ also oscillates at the rate of $\delta f_r$. The frequency oscillation amplitude is proportional to $|m|$. Thus, when measuring the power spectrum of the instantaneous frequency fluctuations (i.e., frequency noise spectrum), sidebands spaced by $\delta f_r$ emerge and their intensities scale as $m^2$. In such a way, we can isolate temporal motion in VFL from amplitude modulation.

Experimentally, we used a pulse shaper (Finisar Waveshaper) to select individual comb lines programmably and employed the delay self-heterodyne method to measure  their frequency noise (Fig. \ref{Fig2}a) \cite{Vahala_OE2022correlated}. Examples of the measured frequency noise spectra are shown in Fig. \ref{Fig2}b. Sharp peaks at $\delta f_r$=1.9 MHz and its harmonics are observed. The intensity of the first harmonics does scale quadratically with $m$ (Fig. \ref{Fig2}c). Therefore, it validates that the multi-line-pair interactions between CP solitons with VFL result in $v_g$ modulation experimentally (see Supplementary Fig. S1 for simulation). 


Our theory and simulation show the pulse motion induced $N$th sidebands have a power extinction ratio with respect to the main line as (Supplementary Sec. 3), 
\begin{equation}
R(m,N) \approx (m\pi f_r \delta t_N)^{-2},
\label{eqnEXR}
\end{equation}
where $\delta t_N$ is the motion amplitude in the $N$th harmonic. To measure these VFL mediated sidebands, we heterodyne beat the microcomb with a continuous wave laser on a BPD. Figure \ref{Fig2}D shows the measured electrical spectrum when beating with the $m$=4 line, which exhibits multiple RF lines spaced $\delta f_r$. The extinction ratio between the main line and the first sideband is about 29 dB. By tuning the wavelength of the laser, we measured the $m^{-2}$ scaling for $R(m,1)$ in Fig. \ref{Fig2}e. The inset confirms that $R(m,2)$ and $R(m,4)$ also follow the $m^{-2}$ scaling. This roll-off is a signature of pulse motion induced sidebands for all combs, regardless of the actual motion trajectory. The motion amplitudes fitted from $R(m,N)$ are $\delta t_1$=33 fs, $\delta t_2$=28 fs, $\delta t_3$=10 fs and $\delta t_4$=25 fs.

\noindent \textbf{Temporal vibration in VFL soliton trimer.} A more direct observation of soliton motion can be measured by a BOC \cite{Vahala_NP2021}. It converts relative pulse delay into a voltage signal with an ultrahigh sensitivity, but requires two inputs with an identical repetition rate. Hence, CP soliton dimers with VFL and nonzero $\delta f_r$ cannot be characterized by a BOC. To resolve this incompatibility, we generated a CP soliton trimer, comprising 1 soliton in one direction and 2 solitons in the other direction. Thus, the 2 solitons within the trimer can be used as the input for the BOC in Fig. \ref{Fig3}a. 

For the 2-soliton within a CP soliton trimer, an example of the measured spectrum is shown in Fig. \ref{Fig3}b. The spectrum exhibits fringes due to the 2-soliton interference. However, different from typical 2-soliton spectra whose fringe contrast is uniform, the contrast of the spectrum in Fig. \ref{Fig3}b decreases with $|m|$ (dots deviate from the fit further with a large $|m|$, see green dots for an example). Such a decrease is evidence of temporal vibration between two pulses \cite{Grelu_OL2006vibrating} (Supplementary Sec. 4).

When sending the 2-soliton stream into the BOC, the output is shown in Fig. \ref{Fig3}c. An oscillation with a peak-to-peak amplitude of about 140 fs was measured. The corresponding motion spectral density is plotted in Fig. \ref{Fig3}d. Note that the BPD within the BOC has a bandwidth of 4 MHz; the high frequency harmonics experience extra signal distortion. 
The measured motion is also well reproduced in our simulation (Supplementary Fig. S4). The complex motion trajectory also suggests the distribution of the Rayleigh backscatterers in the Si$_3$N$_4$ microresonator is highly stochastic and irregular. 


\noindent \textbf{DCS using Si$_3$N$_4$ CP solitons.} With the VFL relevant sidebands, whether Si$_3$N$_4$ CP solitons can be used for precise DCS becomes an open question. To address it, we used an experimental setup shown in Fig. \ref{Fig4}a for DCS. A pulse shaper encoded with sinusoidal- or Lorentzian-absorption was used as virtual samples. The DCS measured transmission with a 0.5 s integration time is shown in Fig. \ref{Fig4}b, in excellent agreement with the applied absorption spectra. The measurement bandwidth spans 40 nm and is limited by the passband of the pulse shaper, instead of the microcomb. 

Microcomb DCS measured with a single-frame interferogram has not been demonstrated, to our knowledge. Our Si$_3$N$_4$ CP soliton system enables this measurement with an acquisition time as short as 1/$\delta f_r$=0.6 $\mu$s (VFL with $n$=32, see Supplementary Fig. S5). Figure \ref{Fig4}C shows the sine-absorption amplitude fitted from DCS data with an averaging time of 0.6 $\mu$s and 5 $\mu$s. The single-frame DCS with 0.6 $\mu$s measurement time already recovers the set absorption amplitude of 0.2, but with a larger standard deviation than the 5 $\mu$s measured data (inset of Fig. \ref{Fig3}c). We further characterize the measurement precision by Allan deviation of the measured sine-absorption amplitude, which reaches 3$\times$10$^{-4}$ at 80 ms and is limited by the measurement time. The CP solitons DCS also shows high spectroscopic signal-to-noise ratio (sSNR, ratio of the spectral line powers to the spectral noise, see refs. \cite{Newbury_Optica2016,ycas2018high} and Supplementary Sec. 5) as plotted in Fig. \ref{Fig4}d. 
When comparing the normalized average sSNR with other reports \cite{Vahala_NC2021architecture,yan2017mid,nader2019infrared,muraviev2018massively,bernhardt2010cavity,ycas2018high}, our Si$_3$N$_4$ CP soliton spectrometer is quite high (see Supplementary Table S1 for a detailed comparison). These measurements confirm the potential of using Si$_3$N$_4$ CP solitons for fast and precise DCS with high sSNR, despite the VFL sidebands. 

We further adjusted the sine-absorption amplitude of the virtual sample and compare it with the sine-fitted amplitude from DCS. Good agreement is observed for relatively low absorption. When the absorption is relatively strong, the measured amplitude is lower than the set absorption due to the VFL sidebands (Figs. \ref{Fig4}E, F and Supplementary Fig. S6). In other words, the VFL limits the measurement dynamic range. By further smoothening the Si$_3$N$_4$ waveguide and lowering the backscattering rate, the dynamic range may be extended (but backscattering is needed for VFL). Silica CP solitons should have larger $R(m,N)$ and a larger dynamic range.

\noindent \textbf{Discussions.} Looking forward, the photonic integrated coherent DCS systems using CP solitons can be realized in the near future (Fig. \ref{Fig4}g). In particular, heterogeneous integration of pump lasers with Si$_3$N$_4$ PICs have been demonstrated \cite{Bowers_Science2021laser}. Electro-optical and acousto-optical control of the pump can be implemented by integrating with the thin film lithium niobate platform \cite{Boes_Science2023lithium,Loncar_AOP2021integrated}. The ultimate portable DCS sensors can be invaluable for precise spectroscopy in cluttered environments or even outer space.

Our work shows that VFL endows high mutual coherence for integrated Si$_3$N$_4$ CP solitons, but the participation of multiple line pairs in VFL causes the solitons to vibrate at a rate of $\delta f_r$. Our measurements reveal the signature of pulse motion induced fine comb line structures, that should exist for other comb generators too. The attainable extinction ratio $R(m,N)$ still allows precise DCS using Si$_3$N$_4$ CP solitons, but it limits the dynamic range. The rhythmic interaction dynamics should be observable for other material platforms and co-pumping configurations \cite{Kippenberg_NP2018spatial}. From the perspective of soliton physics, Si$_3$N$_4$ PICs facilitate the fabrication of coupled microresonators \cite{Vahala_NP2023soliton,Kippenberg_NP2021emergent,VTC_2023surpassing} or even more complex topological structures \cite{Chembo_NP2021topological,Kippenberg_SA2022protected} in a single chip. The synergy of rhythmic interactions within novel PIC designs can pave a pathway towards uncharted territory in soliton research.

\vspace{3 mm}



\vspace{1 mm}

\noindent \textbf{Data Availability}
The data that supports the plots within this paper and other findings are available.

\vspace{1 mm}
\noindent \textbf{Code Availability}
The code that supports findings of this study are available from the corresponding author upon request.

\noindent \textbf{Acknowledgments}
The silicon nitride chip used in this work was fabricated in Qaleido Photonics. This work is supported by the National Key R\&D Program of China (2021YFB2801200), by the National Natural Science Foundation of China (62250071, 62175127), by the Tsinghua-Toyota Joint Research Fund, and by the Tsinghua University Initiative Scientific Research Program (20221080069). J.L. acknowledges support from the National Natural Science Foundation of China (12261131503) and Shenzhen-Hong Kong Cooperation Zone for Technology and Innovation (HZQB-KCZYB2020050).

\vspace{1 mm}
\noindent\textbf{Author Contributions} 
\vspace{1 mm} Z.W. ran the experiments with assistance from Y.W., Y.D., and C.Y. Y.W. ran the simulations and theoretical analysis. B.S., C.S., W.S. and J.L. prepared and characterized the Si$_3$N$_4$ chip. The project was supervised by C.B.   

\noindent \textbf{Competing Interests} The authors declare no competing interests.

\bibliography{main}

\end{document}